# Virtual Reality

**Dan L. Lacrămă, Dorina Fera,**
„Tibiscus" University of Timişoara

ABSTRACT: This paper is focused on the presentation of Virtual Reality principles together with the main implementation methods and techniques. An overview of the main development directions is included.

## 1. Generalities

The concept of virtual reality (VR) refers to a system of principles, methods and techniques used for designing and creating software products in order to be used by help of some multimedia computer systems with specialized peripheral systems.

This provides the possibility to change the way in which the man can perceive the surrounding reality by simulating and modeling an artificial space.

- Competitive multimedia computers have permitted to design and run software programs able to realistically simulate the surrounding reality and to produce a high interactivity level.
- The sophisticated peripheral equipments are permanently developed to variants able to induce to the human operator perceptions that are closer to the natural ones.

All the media replicating the reality are included in the generic category of "virtual reality". The equipments and technologies by which we can interact in a virtual reality are known as VR equipments and VR technologies.

The virtual reality systems can be:
- immersive VR;

137

- simulation VR;
- projected VR;
- telepresence VR;
- augmented reality VR;
- desktop VR.

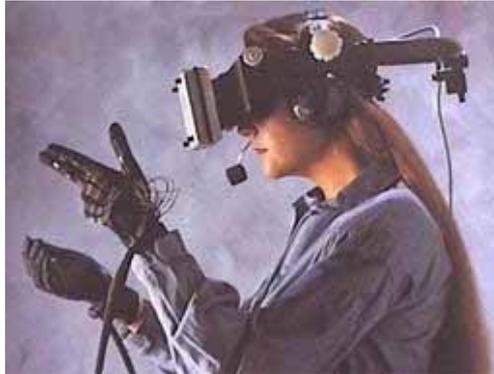

**Figure 1.1** VR Equipment

The main elements in designing and creating the virtual environment are:
- the 3D graphic component by which visualization is provided;
- the model of the virtual environment together with the synthetically elements which compound it;
- the control engine for the motion of the component artificial entities;
- the feedback in real time on the part of both the participants to the interaction man-computer;
- the simulation/control engine of the behavior of the synthetically entities and natural phenomena developed in the simulated environment.

The man's interaction with the environment is done by his senses and limbs. This interaction is bi-directional: **environment → human being**, through the senses, and **human being → environment**, through his limbs. The contact with the reality of the environment is permanent. Yet, at the level of the human being's brain this interaction can be simulated.

The used devices are multimedia peripheral equipments. The figure 1.1 displays such VR equipments. Generally, there are used: 3D glasses, VR

Head mounted display, 3D monitors, VR gloves, steering wheels, game pads able to communicate bidirectional with the virtual environment ("force feedback"), and trackers following the human body's movements.
The most important technological efforts are oriented in order to simulate the 3-dimensional vision and the tactile sense.

## 2. The three-dimensional vision

The 3D vision can be simulated by stereoscopy, meaning the display of two or more images perceived by the brain through the eyes and recompounded by it in a spatial image similar to what we naturally perceive in the everyday reality.

The normal human vision is stereoscopic. Each eye can see one image and by overlapping the two images the human brain receives and processes a final image which is 3D at the level of perception as it is shown in figure 2.1.

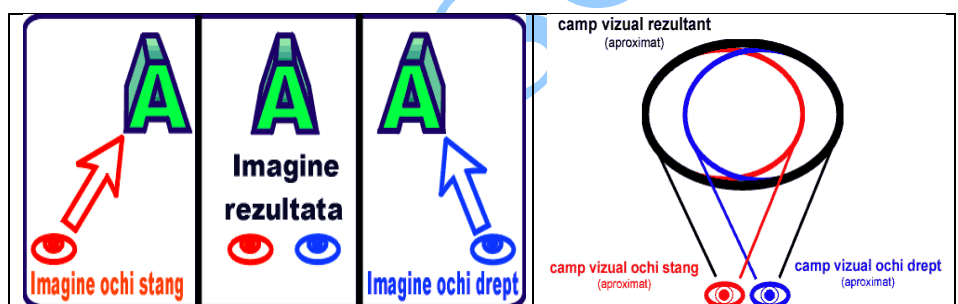

**Figure 2.1.** The human stereoscopic vision

The stereoscopic visualizing equipments and the corresponding technologies must generate a separate image for each eye, having the dimensions as close as possible to the normal visual field and of a high resolution. These images are received by our brain through visual sensors and processed identically with the natural environment.

The 3D visualizing devices most frequently used are:
a) Anaglyphic glasses
b) Shutter glasses;
c) Head Mounted Display

- d) 3D monitors
- e) Volumetric displays
- f) Holographic display systems
- g) Stereoscopic projection systems with DLP projectors.

**a. Anaglyphic glasses** are the simplest devices by which we can see 3D imagine. They have colored lenses which act as color filters (the most frequent are red/blue or red/cyan). They are based on the idea of separating an image in two sub-images: one that takes just the red tints from the initial image and the other taking just the blue + cyan ones. The two images seen from different perspectives (corresponding to the two eyes) overlap, the result being the re-compounding of the initial image which seen through the anaglyphic glasses becomes a three-dimensional image.

**b. The shutter glasses (**for the systems with CRT monitors) have a LCD screen in front of each eye. When the image for the left eye is displayed the right eye is blocked. The on/off switch on each eye is very fast and synchronized with the generation of the corresponding images on the screen, with the result of perceiving a 3D image. Using the 3D glasses in optimal functioning conditions (refresh frequency of 100-180Hz), is identical - from the point of view of the human perception – to watching, without glasses a TV or a monitor with a refresh of 50-90 Hz.

**c. The Head Mounted Display** represents the basic equipment of a virtual reality professional system. The VR Head mounted display has a screen (TFT or LCD) in front of each eye. On these a different image for each eye is displayed, exactly as for the human sight. There are also head mounted displays with incorporated video cameras with the focus in front of each eye. (Trivisio Technologies-Ge). The main parameters: the image resolution (800x600, 1024x768, etc), the field of vision dimensions, (a standard rectangular 4:3 or 16:9 or even a broad field of vision quasi - identical with the human one.)

**d. 3D monitors** are stereoscopic visualizing equipments, relatively recently produced. The main component of such a monitor is an optical device – a matriceal "in waves" selection filter. The screen displays an image that is in fact a combination of more perspectives of the same scene. The used resolution is 1024x768 multiplied by 16, 24 or 32 perspectives and diagonals of 17", 19"or 40". The 3 D monitors are applied in interactive graphic presentations, in commercial campaigns, scientific applications and 3D design programs. The stereoscopic display systems with dimension 1.5-6 m start to find their utilization in many domains in which the three-dimensional visualization with no 3D glasses is needed. The 3D effect is based on the presented technology, but it is temporally prohibitively expensive.

**e. The 3D volumetric displays** contain 20 different LCD's placed one after another and a DLP projector at their back. The light of the projector rapidly scans the LCD panels successively, permitting more than 20 complete refreshes/second for all the 20 LCD's. These spatial devices can display full-color objects occupying a spatial volume, allowing the user to have complete and circular vision (360 degrees). For the static objects in a slow movement the 3D effect is remarkably created. Yet, for those in fast movement, sometimes errors may occur. Additionally, the price of these devices is non-competitive (around 50.000 $). The actual technology is however promising and it continues developing; therefore it could be a good solution in the future.

The holographic display systems obtain 3D images by combining the images for the left eye (L) with those for the right one (R) on a holographic digital display. The system accepts images in stereoscopic formats being also compatible with the 3D graphic software. These holographic devices are more and more used in advertising. The holograms are not a technological novelty but the total holographic images: 360 degrees visible, color, high resolution, moving and floating in space have become possible only recently.

**f. Stereoscopic projection systems** allow the display of the stereoscopic images through the DLP video-projectors. One of the following stereoscopic solutions can be used:
- Vertical and horizontal polarization of the L and R images done by the two DLP projectors, separated for the two images corresponding to the left, respectively right eye. The visualization is done through glasses having linearly or circularly polarized lenses.
- Page Flipping with shot delay, meant to prevent the "Ghosting" phenomenon usually present with the CRT monitors. The visualization is done by 3D shutter glasses.

Although some of these equipments are very sophisticated and frequently very expensive, they are temporally not able to provide a perfect 100% simulation of the natural 3D sight. O series of technical limitations determine either representing errors of the simulated objects or the appearance of some undesired ghosts.

141

## 3. Tactile sense

**The haptic is** the scientific domain that studies the simulation of the pressure, texture, temperature, vibration and other sensations perceived by the tactile sense. It is a developing technology. The available systems are based on devices in which the fingers are introduced. The pressure-connected sensors allow the user to "feel" the images on the screen, to explore them, to feel them cold or warm, hard or soft, rough or smooth.

The complete simulation of the tactile sense supposes that the user has the feeling that he/she can catch and manipulate the objects in the virtual reality. This involves not only creating the apprehension but also modifying the position of the objects by direct manipulation.

Simulating the tactile sense adds interesting applications in domain such as:
- Industrial design;
- Simulators for training the pilots, astronauts, drivers, sportspeople, etc;
- Education;
- Medicine;
- Virtual scientific experiments;
- Computer games;
- Entertainment.

An interesting application would be that of virtually creating and manipulating on the computer some impossible objects such as the "Moebus' Band".

The best-known tactile simulation device is the **Sensorial or data glove** presented in figure 3.1. This device has for each finger and for each palm, small vibrato-tactile simulators. Each simulator can be individually programmed to offer the sensation of touching.

The simulators can generate a whole area of pulsation or sustained vibrations sensations. By combining them, a complex tactile feedback pattern can be generated. The user has the possibility of creating his profile by himself. There is also the possibility of introducing in the material of the glove some conductor plates that have the property of warming or cooling up according to the polarity applied at the input electrodes.

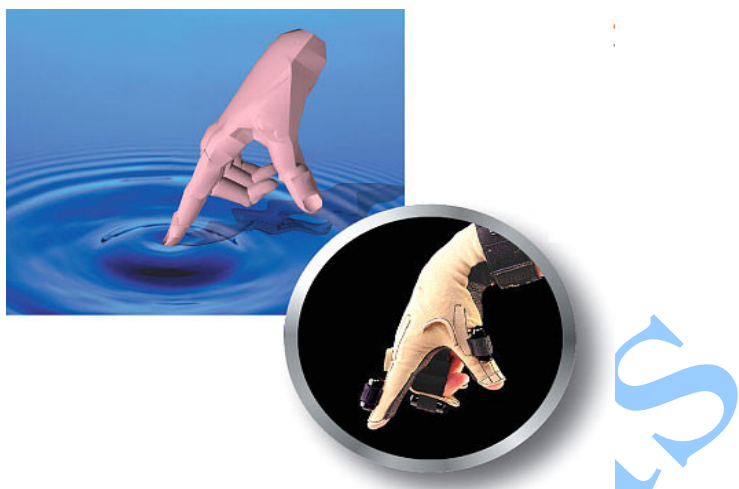

**Figure 3.1.** Sensorial glove

## 4. Descriptive files for the VR worlds

**VRLM –** Virtual Reality Modeling Language - usually pronounced *vermal* – is a standard format file for representing the 3D interactive, vectorial graphics, especially conceived for the World Wide Web. A file based on VRLM is usually a text file, in which the characteristics of the virtual elements are described. For example, for a 3D polygon both the features referring at its position in space, and those referring to its color, texture, can be indicated.
In addition, animation, sounds, representations of light sources and other virtual aspects can be added to interact with the user or they can be induced by an external event (for instance by a timer).
     VRLM files are currently named "worlds" and they have the. wrl extension. They are generally compressed to be more rapidly transferred on Internet. Most of the 3 D modeling programs can save objects and scenes in VRLM format. In order to monitor the collective development of this standard and to ensure the compatibility between the versions the Web3D Consort was made up. The first VRLM version appeared in November 1994. The last official version was VRLM97, also used on some 3D chat sites. Eventually VRLM was excelled by X3D.

The extensible 3D (X3D) is a standard conceived for the 3 D content delivery and representation. It combines both the geometric proportions of the objects and their behavioral descriptions (the way an object behave during running) in one file that may have different formats, including XML.

It is also the next step after VRLM and its architecture incorporates both the most recent developments in the domain of the graphic representation hardware as well as the improvements based on the years of experience of the community that has developed VRML.

## 5. Conclusions

The Virtual Reality is a recent research domain, but the progresses made in the last few years by the companies developing VR systems prove that in the next future this technology will be high-quality and cheap enough to become common good.

VR's applications seem to grow in number and diversify covering very different domains from research to entertainment. Thus the interest in its development is quite high.